\documentclass[twocolumn,english,prl,superscriptaddress,reprint,showpacs,showkeys]{revtex4-1}

\usepackage[T1]{fontenc}
\usepackage{geometry}
\usepackage{float}
\usepackage{booktabs}
\usepackage{amsmath}
\usepackage{amssymb}
\usepackage{graphicx}
\usepackage{setspace}
\usepackage[utf8]{inputenc}
\usepackage{textcomp}

\usepackage{accents}

\makeatletter

\usepackage[newcommands]{ragged2e}

\usepackage[unicode=true,pdfusetitle,
 bookmarks=true,bookmarksnumbered=false,bookmarksopen=false,
 breaklinks=false,pdfborder={0 0 1},backref=false,colorlinks=false]
 {hyperref}
\hypersetup{
 colorlinks,linkcolor=blue,citecolor=blue,urlcolor=blue}

\usepackage{xcolor}

\usepackage{babel}
\usepackage{dsfont}
\begin{document}
\title{Microscopic theory of the inverse spin galvanic  effect in anisotropic Rashba models}
\author{Alessandro Veneri}
\affiliation{Dipartimento di Ingegneria dell'Informazione, Elettronica e Telecomunicazioni, Universit\`a  La Sapienza,  Via Eudossiana, 18, 00184 Roma, Italy}
\affiliation{Dipartimento di Matematica e Fisica,  Universit\`a  Roma Tre, Via della Vasca Navale 84, 00146 Roma, Italy}
\author{Francesco Quintavalle}
\affiliation{Dipartimento di Matematica e Fisica,  Universit\`a  Roma Tre, Via della Vasca Navale 84, 00146 Roma, Italy}
\author{Thierry Valet}
\affiliation{MPhysX O\"U, Harju maakond, Tallinn, Lasnam\"ae linnaosa, Sepapaja tn 6, 15551, Estonia}
\author{Roberto Raimondi}
\affiliation{Dipartimento di Matematica e Fisica,  Universit\`a  Roma Tre, Via della Vasca Navale 84, 00146 Roma, Italy}

\begin{abstract}
The Rashba spin-orbit coupling (SOC) is a well-known mechanism for the spin-charge interconversion via the inverse and direct spin galvanic effects. The lack of a full inversion symmetry allows the coupling of the charge current and spin density. In this paper we investigate this phenomenon when the in-plane rotational symmetry is lowered to the $C_{2v}$ and $C_{3v}$ symmetry groups, whereby the electron spectrum becomes anisotropic. We find that in the $C_{2v}$ case, depending on the ratio between the Rashba SOC strengths along the principal axes, the non-equilibrium spin density deviates notably from the $90^o$ degrees rotation, with respect to the applied electric field, familiar in the isotropic case. In the $C_{3v}$ case, when a warping cubic-in-momentum term is present, whereas the standard $90^o$ degrees rotation of the spin density remains, the spin-charge interconversion depends on the intensity of the warping itself.
The microscopic theory takes into account disorder including vertex corrections, both via the diagrammatic implementation of the Kubo formula and via the quantum kinetic theory.  We show that vertex corrections are crucial to capture the details of the inverse spin galvanic effect in contrast to previous treatments based on the constant broadening approximation.
\end{abstract}
 
\maketitle

\section{Introduction}
Spin-dependent phenomena in nonmagnetic materials lacking inversion symmetry have attracted a lot of attention in the last few decades, as these systems enable all-electrical control of the spin degree of freedom (DOF) without the need for external magnetic fields \cite{datta_1990, Nitta_97, Silsbee04, Hirohata_2014, Ganichev_2016, Soumyanarayanan2016}. The underlying mechanisms arise from the relativistic spin-orbit coupling (SOC), which induces a spin-momentum locking that renders the spin and momentum DOFs fully interdependent \cite{Dresselhaus_1955, Rashba_1960,Ivchenko1978,Vorobev1979, Kane_2005,Dyakonovbook08}. Sources of SOC can manifest both uniformly and locally, with the former being induced by the system's geometry and the associated crystalline potential \cite{Kochan_2017, Cysne_2018} while the latter originates from randomly distributed fluctuations or proximity to adatom impurities \cite{Neto_2009, Weeks_2011, balakrishnan_2014}.\\
Promising for the development of experimental spintronics devices \cite{Liu_2004}, SOC landscapes enable charge-spin conversion phenomena that generate spin currents and densities.
In particular, two effects, together with their Onsager reciprocal counterpart, emerged as the most intriguing: the spin-Hall effect (SHE) \cite{DPshe71, Sinova_2004, Murakami2004,sinova2015,Perkins_2024} and the inverse spin-Galvanic effect (ISGE) \cite{aronov_1989, Ivchenko90,Ganichev02,Shen_2014, Gorini_2017, Offidani_2017, Veneri_2022}, also called Edelstein effect (EE). 
The former converts a charge current, driven by an external electric field, into a transverse spin current, while the latter converts an electric current into a non-equilibrium spin polarization. Their origin can be traced either to the spin-dependent scattering of electrons by randomly distributed impurities, referred to as extrinsic phenomena \cite{engel2006,tse2006,hankiewicz2008,Raimondi_2009,Raimondi_2012}, or to the relativistic electronic structure of the material, where disorder plays a secondary role, referred to as intrinsic phenomena. For example, graphene exhibits an extrinsic SHE in the presence of local spin-orbit interaction via the skew scattering mechanism \cite{Milletari_2016} and a quantized intrinsic spin-Hall conductivity associated with the $Z_2$ topological invariant \cite{Kane_2005_B}. The same reasoning applies to the EE, which can either be generated by random sources of SOC via the anisotropic spin-precession mechanism \cite{Huang_2016}, or by the uniform Rashba SOC \cite{bychkov_1984}.\\
The latter has served as a paradigmatic model system for studying the EE and has undergone extensive theoretical scrutinity, leading to significant developments over the years, although the EE itself has received less attention compared to the SHE. After the earliest results revealed the possibility of generating an in-plane spin polarization perpendicular to an applied static voltage in two-dimensional electron gases (2DEG) \cite{kato2004,sih2005,rojas2013}, successively also obserbed in non-magnetic interfaces\cite{rojas2013}, LAO/STO systems\cite{Lesne2016} and topological insulators\cite{Shiomi2014,Mellnik2014}, a space-time dependent drift-diffusion theory has been formulated \cite{Burkov_2004, Gorini_2010, Gorini_2012}. This framework was later extended to surface states of topological insulators \cite{Burkov_2010}, engineered graphene systems \cite{Ferreira_2021}, and van der Waals heterostructures \cite{Veneri_2022_B}, the latter revealing the possibility of generating twist-angle controlled spin polarization at arbitrary angle with respect to the applied field. \\
This paper focuses on a generalization of the traditional disordered Rashba model accounting for anisotropies in the effective mass and the linear Rashba parameter, as well as third order warping terms, predicted to arise in the presence of $C_{3v}$ symmetry \cite{fu2009,Frantzeskakis_2011}. This analysis is needed for a more realistic understanding of the EE in 2DEGs, as crystalline field effects can modify the simple isotropic free-electron kinetic Hamiltonian, especially away from the high point-group symmetry surfaces \cite{Simon_2010}. 
Although we will not discuss in the present paper the microscopic details in real materials, 
it is worth mentioning that the description of spin splitting in surface states may be quite complex and may involve furher effects besides the Rashba SOC \cite{ibanez2013,krasovskii2014}.\\
The problem of an anisotropic Rashba SOC has been addressed only rarely and without comprehensive detail in the literature \cite{Johansson_2016, miatka2019, Montecinos_2023, Gaiardoni_2025}, mostly focusing on the dependence on the Fermi energy and confining to the relaxation time approximation (RTA) \cite{grosso_2000}, where transport lifetimes are replaced by mere constant values.
In this regard, linear response theory is the main formalism adopted to investigate such transport phenomena. It is typically reduced to two main techniques: the Boltzmann equation (BE) \cite{ziman_2001}, with its extensions \cite{mishchenko2004, khaetskii2006,Xiao_2017, Sekine_2017}, and the Kubo formula \cite{Kubo_1956, Kubo_1957,BASTIN_1971, Streda_1975}. The former is strictly valid in the dilute limit, which provides the leading contribution to the system's response, while the second provides a complete quantum-mechanical description of the system, incorporating geometrical Berry phase effects \cite{Xiao_2010} and impurity-related quantum processes \cite{Ado_2015,ado_2017}. In the semi-classical limit, that is the regime considered in this work, there exists an exact correspondence between the two frameworks \cite{Sinitsyn_2007,Valet_2025_theory}: the self-consistent solution of the BE, which determines the transport lifetimes describing the out-of-equilibrium distortion of the distribution function, translates into the vertex renormalization in the Kubo formalism, together with the proper evaluation of the Green's function (GF) self-energy. Due to the challenging calculations required for a consistent treatment of disorder, the BE is often solved within the RTA. This corresponds to the constant broadening approximation (CBA) in the Kubo formalism \cite{Sinitsyn_2007_B, Go_2024,ovalle_2023}, in which vertex corrections are totally ignored. However, this uncontrolled approximation is often seriously problematic. A well-known example is the SHE in 2DEGs: while neglecting the vertex corrections predicts a finite spin-Hall conductivity, proper calculations reveal a vanishing response \cite{Raimondi_2005,Inoue_2004,dimitrova2005,chalaev2005}.\\
This paper represents another such case. We will show that the contribution of mass anisotropies to the EE cancels exactly once vertex corrections are taken into account, opposite to what is currently predicted by the literature \cite{Gaiardoni_2025}. In contrast, Rashba anisotropies make the EE itself anisotropic, influencing both its magnitude and orientation depending on the direction of the applied electric field. As a result, since the direction of the magnetization is generally no longer perpendicular to the electric current, we propose leveraging this deviation for direct experimental measurements of Rashba anisotropies. Our analysis employs both the Kubo formalism and the BE approach, with the former yielding analytical results and the latter numerical ones.\\
The remainder of this paper is organized as follows. In Sec. \ref{sec:method} we present the model Hamiltonian, briefly review the Kubo formalism used for the analytic calculations, and describe the numerical implementation of the BE. Moreover, here we clarify the connection between the RTA and the CBA. Sec. \ref{sec:res} is devoted to the calculation of the spin susceptibility. We first reproduce the literature results for mass and Rashba anisotropies independently within the CBA. Then, we extend the analysis by including the vertex corrections and the microscopic parameter-dependent self-energy in the Kubo formalism, coupled with the numerical solution of the BE beyond RTA. Finally, we investigate the effect of anisotropic warping in systems with $C_{3v}$ symmetry. In Sec. \ref{sec:conc} we present our conclusions.

\section{Model system and Methodology}\label{sec:method}

\begin{figure}
	\begin{centering}
		\includegraphics[scale=0.45]{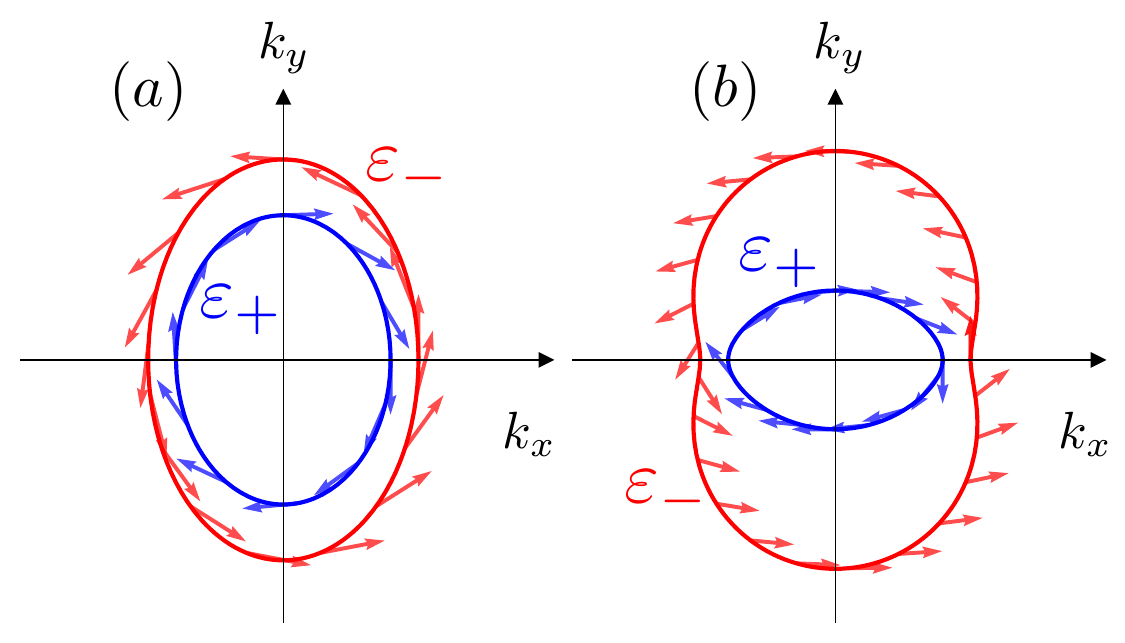}\caption{Fermi surfaces obtained from Eq. (\ref{Ham}) in regime II with anisotropic mass (a) and spin (b) terms. The lower and upper bands, corresponding to the outer (red) and inner (blue) curves, are labeled $\varepsilon_-$ and $\varepsilon_+$, respectively. The direction of the arrows indicate the spin expectation values. Mass anisotropy is characterized by the ratio $m_y/m_x=2$, while the Rashba coupling anisotropy is defined by $\alpha_y/\alpha_x = 5$. The reference system is the surface state of Au(1 1 1), with effective mass $m=0.27\,m_{\mathrm{e}}$, with $\mathrm{m_e}$ being the electron mass, $\alpha = 0.33\,\mathrm{eV}\,\text{\AA}$, and Fermi energy $\varepsilon_{\mathrm{f}}=0.475$\,$\mathrm{eV}$ \cite{Cercellier_2006}. For a better representation of the Fermi surfaces, we increased the Rashba coupling by a factor of 2. The Rashba splitting is $\sim 0.12\,\mathrm{eV}$.}\label{fermi_disks}
		\par\end{centering}
\end{figure}

The anisotropic Rashba model for 2DEG systems with $C_{2v}$ symmetry  can be described by the low-energy Hamiltonian \cite{Simon_2010}
\begin{equation}
    H = \frac{p_x^2}{2m_x} + \frac{p_y^2}{2m_y} + \alpha_y p_y s_x - \alpha_x p_x s_y, \label{Ham}
\end{equation}
where $m_{x(y)}$ is the effective mass along the x(y) axis and $s_i$ is a Pauli matrix $i$. Due to Rashba splitting, spin degeneracy is lifted, and the energy eigenvalues form two shifted parabolas, distorted by both mass and Rashba anisotropies
\begin{equation}
\varepsilon_\pm (p_x,p_y)=\frac{p_x^2}{2m_x} + \frac{p_y^2}{2m_y} \pm
\sqrt{(\alpha_y p_x)^2+(\alpha_x p_y)^2}.
    \label{eigenvalues}
\end{equation}
Analogously, we can define two distinct energy branches corresponding to an upper and lower band. The Fermi energy can then be positioned in such a way to intersect either a single branch (regime I) or both branches (regime II). In what follows, we will focus on the latter case, where the Fermi surface consists of two deformed disks with spin locked and perpendicular to the electron momentum. In this regime, the spin textures of the lower and upper bands wind in opposite directions, anticlockwise and clockwise, respectively (see Fig. (\ref{fermi_disks})).\\
Eq. (\ref{Ham}) is the starting point for our linear response calculations, whose technical details will be presented below. Assuming a slowly-varying electric field $\mathbf{E}$ in both space and time, the spin density response takes the form 
\begin{equation}
    S_{\alpha} = K_{\alpha \beta}E_{\beta}, \label{basic_LR}
\end{equation}
where $K_{\alpha \beta}$ is the $3\times2$ spin susceptibility tensor, $\alpha=x,y,z$, $\beta=x,y$, and the Einstein summation rule is assumed. 

\vspace{0.1cm}

\textit{The Kubo formula and diagrammatic theory--} Within the Kubo formalism, the disorder-averaged spin-current response can be separated into two components, Fermi sea and a Fermi surface \cite{Bonbien_2020}. Restricting our analysis to the dilute limit, wherein the impurity concentration $n_i\ll1$, we can extract the leading contribution to the spin susceptibility $\sim1/n_i$ by safely neglecting the Fermi sea term and products of the GF belonging to the same sector, i.e., retarded or advanced,
\cite{Milletari_2016}, finally obtaining the Kubo-Streda formula \cite{Smrcka_1977, crepieux_2001}
\begin{equation}
    K_{\alpha \beta}=\frac{1}{2\pi}\int \frac{d\mathbf{p}}{(2 \pi)^2}\,\mathrm{tr}\langle\left[\frac{s_\alpha}{2} G^{\mathrm{R}}_{\mathbf{p}}j_{\beta}G^{\mathrm{A}}_{\mathbf{p}} \right]\rangle_{\mathrm{dis}}\label{Kubo_1}
\end{equation}
where the trace is over the internal matrix indices, $j_{\beta}=\partial H/\partial p_{\beta}$ is the electric current operator, i.e., here defined without the electric charge $-e$, $G^{\mathrm{R(A)}}_{\mathbf{p}} = 1/(\varepsilon-H\pm i0^{+})$ is the clean retarded (advanced) GF, and we made use of natural units $\hbar \equiv 1 \equiv e$. \\

\begin{figure}
	\begin{centering}
		\includegraphics[scale=0.38]{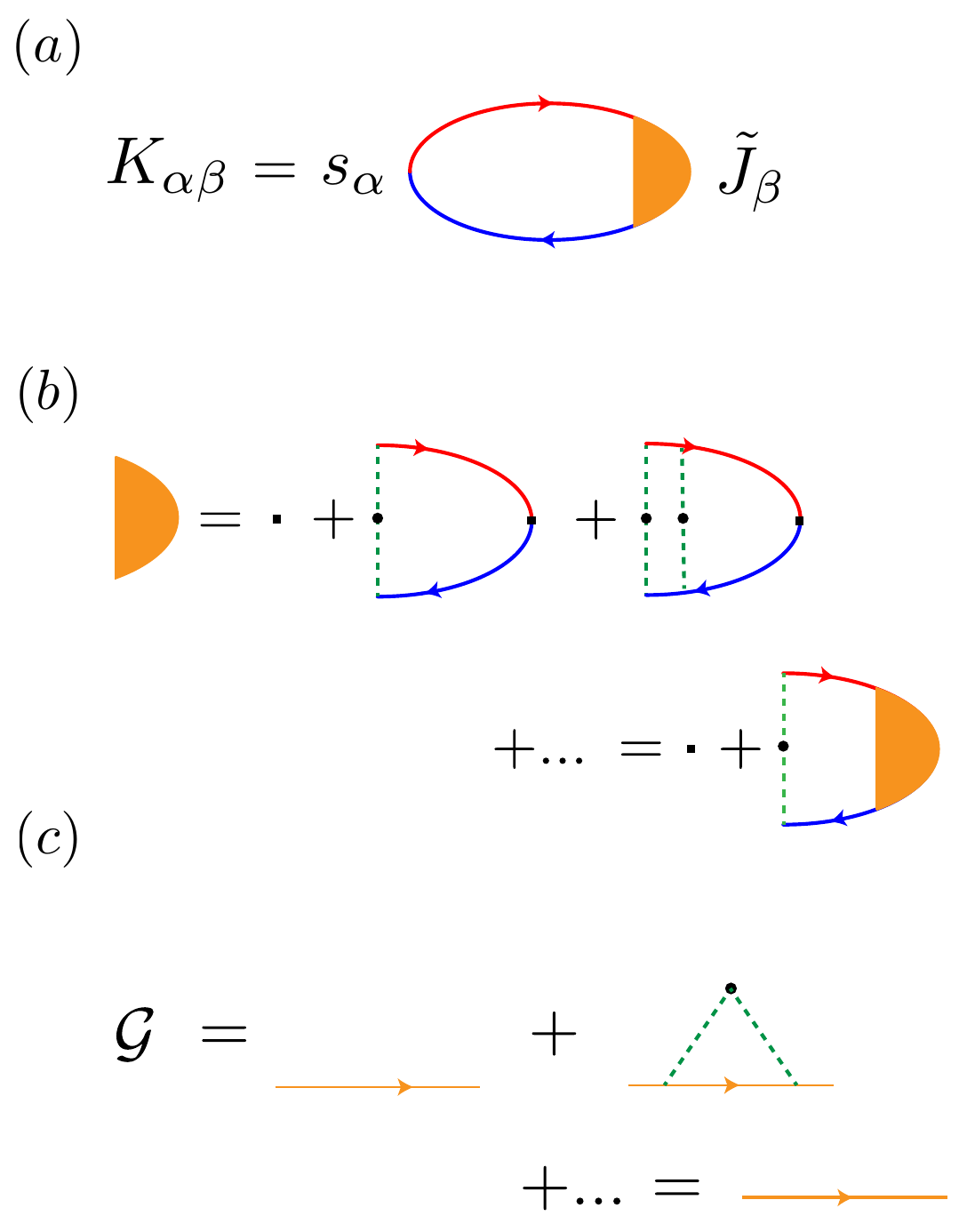}\caption{Diagrammatic representation of the spin-current response function. The impurity-averaged bubble in (a) depicts in the Kubo formula, where blue and red solid lines with arrows represent the advanced and retarded GF sectors, respectively. The dressed vertex is indicated by orange shading. Its skeleton expansion is shown in (b), where the dashed green lines denote the averaged interaction with a single impurity, represented by a black dot. The bare current vertex is identified by a black square. (c) Expansion of the GF within the FBA, where thin lines denote bare GFs and thick lines denote dressed GFs.}\label{diagrammatics}
		\par\end{centering}
\end{figure}

Disorder averaging, here assumed to be within the first Born approximation (FBA) \cite{Sinitsyn_2007_B}, is denoted by $\langle...\rangle_{\mathrm{dis}}$.
Using the standard rules of diagrammatics \cite{Di_Castro_Raimondi_2015} (or functional approaches to nonequilibrium field-theory methods \cite{Kamenev_2011}), Eq. (\ref{Kubo_1}) becomes 
\begin{equation}
    K_{\alpha \beta}=\frac{1}{2\pi}\int \frac{d\mathbf{p}}{(2\pi)^2}\,\mathrm{tr}\left[\frac{s_\alpha}{2} \mathcal{G}^{\mathrm{R}}_{\mathbf{p}}\tilde{J}_{\beta}\mathcal{G}^{\mathrm{A}}_{\mathbf{p}} \right],\label{Kubo_2}
\end{equation}
depicted in Fig.(\ref{diagrammatics}a) as a dressed bubble, where $\tilde{J}_{\beta}$ is the renormalized current vertex and 
\begin{equation}  \mathcal{G}^{\mathrm{R(A)}}_{\mathbf{p}} = \frac{1}{\varepsilon-H-\Sigma^\mathrm{R(A)}} \label{GF}
\end{equation}
is the disorder-averaged retarded (advanced) GF. Under the FBA, the leading contribution to the self-energy in impurity concentration is given by $\Sigma^\mathrm{R(A)}=n_i u_0^2 g_0^\mathrm{R(A)}$, with $g_0^\mathrm{R(A)}$ being the momentum-integrated clean GF and $u_0$ the potential of a single impurity, see Fig.(\ref{diagrammatics}c). Consistently, the renormalized vertex is given by the Bethe-Salpeter integral equation \cite{raimondi_2001,schwab_2002,Raimondi_2005,Offidani_2018}
\begin{equation}
    \tilde{J}_{\beta} = j_{\beta} + n_i u_0^2 \int \frac{d\mathbf{p}}{(2 \pi)^2} \mathcal{G}^{\mathrm{R}}_{\mathbf{p}}\tilde{J}_{\beta}\mathcal{G}^{\mathrm{A}}_{\mathbf{p}},\label{BS_1}
\end{equation}
corresponding to the non-crossing ladder series of impurity scattering, illustrated in Fig.(\ref{diagrammatics}b). 
A convenient way to solve Eq.(\ref{BS_1}) is to separate the renormalized vertex into the bare, momentum-\textit{dependent} term $j_{\beta}$ and the momentum-\textit{independent} correction $\delta J_{\beta}$, such that $\tilde{J}_{\beta} = j_{\beta} + \delta J_{\beta}$ \cite{Veneri_2022}. The result is a Bethe-Salpeter equation for $\delta J_{\beta}$,
\begin{equation}
    \delta J_{\beta} = \delta \bar{J}_{\beta} + n_i u_0^2 \int \frac{d\mathbf{p}}{(2 \pi)^2} \mathcal{G}^{\mathrm{R}}_{\mathbf{p}}\delta J_{\beta}\mathcal{G}^{\mathrm{A}}_{\mathbf{p}},\label{BS_2}
\end{equation}
where $\delta \bar{J}_{\beta}=n_i u_0^2 \sum_{\mathbf{p}} \mathcal{G}^{\mathrm{R}}_{\mathbf{p}} j_{\beta}\mathcal{G}^{\mathrm{A}}_{\mathbf{p}}$ is the first order correction to the current operator. Since the displacement of the current operator in Eq.(\ref{BS_2}) is momentum-independent, it can be factored out, allowing the equation to be solved explicitly, although this is generally a difficult task \cite{Offidani_2018}. The dressed vertex  is obtained as
\begin{equation}
    \delta J_{\beta}=\mathcal{D}_{\beta \nu } \delta \bar{J}_{\nu},
\end{equation}
where 
\begin{equation}
\mathcal{D}^{-1}_{\mu \nu }=\delta_{\mu\nu}-\frac{n_i u_0^2}{2}\mathcal{N}_{\mu \nu }\label{closed_J}
\end{equation}
is called \textit{diffuson} and $\mathcal{N}_{\mu \nu } = \int \frac{d\mathbf{p}}{(2 \pi)^2}\mathrm{tr}\left[s_{\mu}\mathcal{G}^{\mathrm{R}}_{\mathbf{p}}s_{\nu}\mathcal{G}^{\mathrm{A}}_{\mathbf{p}} \right]$.
Once Eq.(\ref{closed_J}) is evaluated, the dressed current vertex is fully determined, and the Kubo formula is thereby completely specified. This approach conveniently eliminates the need for any ansatz for $\tilde{J}_{\beta}$, making the method a transparent and general algorithm for obtaining analytical solutions to transport problems. However, implementing it is far from being easy, especially in the presence of band anisotropies, which make momentum integrations nontrivial.

\vspace{0.1cm}

\textit{Quantum Kinetic theory (QKT)--} A recently developed QKT approach\cite{valet:2023,Valet_2025_theory}, based on the Keldysh-Larkin-Ovchinnikov technique \cite{Keldysh, larkin1975nonlinear}, enables a rigorous, systematic inclusion of impurity-induced semiclassical mechanisms, quantum corrections, and topological effects within the linear response theory, with no need for explicit diagrammatic expansions, but solely relying on the self-energy expression \cite{Valet_2025_theory}. This method has the advantage of providing a clearer physical understanding of the problem under investigation, enabling more streamlined practical calculations, establishing a formal connection with the Kubo formalism, and incorporating macroscopic finite size effects \cite{valet2025}.

In the present case, restricting our interest to the system semiclassical steady-state response, the method translates into a set of $N_{\mathrm{b}}$ linear Boltzmann equations (BEs),
\begin{equation}
    \mathbf{E}\cdot \partial_{\mathbf{p}}\varepsilon_{\mathbf{p},n}
\,\partial_{\varepsilon}f_{\mathbf{p}, n}^{(\mathrm{eq})} = I_n, \label{BE}
\end{equation}
where $N_{\mathrm{b}}$ is the number of bands in the system, $n$ and $m$ are band indices, $\varepsilon_{\mathbf{p},n}$ denotes the eigenvalue in momentum space of the $n$th band, and $f_{\mathbf{p},n}^{(\mathrm{eq})}$ is the equilibrium Fermi-Dirac distribution. The collisional integral $I_n$, which arises from the disorder-induced self-energy for the Keldysh Green function, takes the form 
\begin{equation}
    I_n=\sum_m \int \frac{d\mathbf{p}'}{(2 \pi)^2}\delta(\varepsilon_{\mathbf{p'},m}-\varepsilon_{\mathbf{p},n})\left(\delta f_{\mathbf{p'},m}-\delta f_{\mathbf{p},n} \right)W_{nm} \label{CI}
\end{equation}
with 
\begin{equation}
    W_{nm} = 2 \pi n_i u_0^2\mathrm{tr}\left[\hat{P}_{\mathbf{p},n}\hat{P}_{\mathbf{p'},m} \right]
\end{equation}
being the scattering kernel, $\hat{P}_{\mathbf{p},n}$ the projector into the eigenspace identified by $\varepsilon_{\mathbf{p},n}$, and $\delta f_{\mathbf{p},n}$ the linear displacement of the occupation function of the $n$th band, i.e., $f_{\mathbf{p},n} = f_{\mathbf{p},n}^{\mathrm{(eq)}} + \delta f_{\mathbf{p},n}$.
Once Eq. (\ref{BE}) is solved, the observable of interest, which, in our case, corresponds to the spin density, is calculated, at leading order $\sim 1/n_i$, as 
\begin{equation}
    S_{\alpha}= \sum_n\int \frac{d\mathbf{p}}{(2 \pi)^2} \mathrm{tr}\left[\left(f_{\mathbf{p},n}  \hat{P}_{\mathbf{p},n}\right)\frac{s_{\alpha}}{2}\right], \label{BE_solution}
\end{equation}
which takes the form of Eq. (\ref{basic_LR}). As shown in Ref.\cite{Valet_2025_theory} sub-leading corrections $\sim (n_i)^{0}$ can also be obtained by evaluating the off-diagonal, in the energy basis, matrix elements of the density matrix. However, there is no need to consider them for the evaluation of the spin response.

In concrete analytic calculations, the system of BEs can be solved in two ways: i) by making an ansatz for the out-of-equilibrium distribution function, or ii) by solving the equations iteratively in a Bethe-Salpeter fashion. 
Often, the first-order  correction suggests the suitable ansatz.
The former method ideally closes the system of equations, making them solvable \cite{Milletari_2016}, while the latter is traditionally approached using the RTA. This commonly adopted procedure solves Eq. (\ref{BE}) at first order -- thus neglecting the contribution $\sim \delta f_{\mathbf{p'},m}$ in Eq. (\ref{CI}), often referred to as \textit{scatter-in} term \cite{grosso_2000}-- and replaces the resulting integral with a constant transport time $1/\tau$, i.e.,  $I_n \rightarrow -\delta f_{\mathbf{p},n}/\tau$. The combination of these crude assumptions can be problematic, as they lose details of the scattering mechanisms in play, which depend on both the bare system Hamiltonian and the impurity structure. 
The early success of the RTA is a result of its application to specific, paradigmatic systems, e.g.,  single-band Hamiltonians with scalar electrostatic impurities. In such cases, the linear displacement of the occupation function is proportional to the scalar product between the group velocity, $\partial_{\mathbf{p}}\varepsilon_{\mathbf{p},n}=\mathbf{v}_n$, and the electric field, i.e., $\sim\cos \theta$, while the scattering kernel is momentum-independent. Consequently, the momentum integration defining scatter-in term in Eq. (\ref{CI}) vanishes, thereby validating the RTA. However, its validity is compromised in multi-band systems, such is in the present case, where the influence of SOC during scattering events renders the scattering kernel momentum-dependent and the scatter-in term non-vanishing.

Our QKT environment allows us to frame both approximations within the Kubo formalism. According to Ref. \cite{Valet_2025_theory}, the out-of-equilibrium occupation function at the leading order in the impurity concentration takes the form
\begin{equation}
    f_{\mathbf{p}, n}=-\partial_{\varepsilon}f_{\mathbf{p},n}^{(\mathrm{eq})}\tau_n \sum_i \mathbf{E} \cdot \tilde{\mathbf{J}}_n, \label{Kubo-Bol}
\end{equation}
where $\mathrm{i}/2\tau_n = \textrm{Im}\Sigma^{\mathrm{A}}_n$,  with $\Sigma^{\mathrm{A}}_n$ and $\tilde{\mathbf{J}}_n$ being, respectively, the advanced GF self-energy and the renormalized vertex shown in Eqs. (\ref{GF}) and (\ref{BS_1}), projected on the $n$th band. This expression can be directly compared with the RTA solution of the BE:
\begin{equation}
    f_{\mathbf{p}, n}^{\mathrm{RTA}}=-\partial_{\varepsilon}f_{\mathbf{p},n}^{(\mathrm{eq})}\tau \sum_i \mathbf{E}\cdot \mathbf{j}_n, \label{RTA_sol}
\end{equation}
clarifying the meaning of the RTA within the diagrammatic language: it corresponds to neglecting vertex renormalization in the Bethe-Salpeter equation and replacing the self-energy with a constant independent of the Hamiltonian and impurity structure. The former approximation, as extensively discussed in the literature, has proved inadequate for Hamiltonians with nontrivial structures in spin space \cite{Raimondi_2005}, while the latter effectively neglects the shape of the density of states (DOS), which is particularly questionable here, since the DOS encodes information on the anisotropy.

The need of surpassing the RTA motivated us to develop a numerical algorithm for the exact solution of Eq. (\ref{BE}). In contrast to the Kubo-Streda formula, where the GFs introduce discontinuities in momentum space tricky to handle, this approach is particularly suitable for numerical integrations, thanks to the Fermi surfaces sole contribution to the linear response within the BEs. In particular, the occupation number takes the form 
\begin{equation}
    \delta f_{\mathbf{p}, n}=-\delta (\varepsilon_{\mathbf{p},n}-\varepsilon_{\mathrm{F}})  g_{\mathbf{p}, n}, \label{ansatz}
\end{equation}
with $\varepsilon_{\mathrm{F}}$ being the Fermi energy.
While analytical approaches typically need an ansatz for the angular dependence of the distribution function \cite{Milletari_2016, Huang_2016,Raimondi_2025_application}, Eq. (\ref{ansatz}) shows the most general structure for the occupation number, as the coefficients $g_{\mathbf{p}, n}$ retain full information of both the absolute value of the momentum and its direction. Inserting  our ansatz in Eq. (\ref{CI}), the BE for band $n$ takes the form
\begin{equation}
    \mathbf{E}\cdot\mathbf{v}_n(\phi)=\sum_m \int \frac{d\phi'\,\mathrm{k}_{m}}{(2 \pi)^2}\left(\frac{g_{m}}{\mathrm{v}_{m}}(\phi')-\frac{g_{n}}{\mathrm{v}_{n}}(\phi) \right)W_{nm}(\phi,\phi'), \label{BE_ANG}
\end{equation}
where all quantities are evaluated at Fermi energy; e.g., $\mathrm{k}_{m}$ and $\mathrm{v}_{n}$ are, respectively, the absolute value of the Fermi momentum in band $m$ and the group velocity for band $n$ at the Fermi level. We now discretize the Fermi surfaces in $N_\theta$ angular steps and, accordingly, all the elements in Eq. (\ref{BE_ANG}), with the aim of rewriting the right-hand side of the BE as a Riemann sum. In matrix form, Eq. (\ref{BE_ANG}) can finally be written as \cite{Veneri_Borg_2025}
\begin{equation}
    \mathbf{E}\cdot \bar{\mathbf{v}} = \hat{\mathrm{S}}\, \bar{g},\label{numerical_BE}
\end{equation}
where $\bar{g}$ and each component of $\bar{\mathbf{v}}$ are column vectors of dimension $N = N_{\mathrm{b}} \times N_{\theta}$, and  $\hat{\mathrm{S}}$ is the collisional integral expressed as an $N\times N$ matrix. The occupation number is then obtained as
\begin{equation}
    \bar{g} = \hat{\mathrm{S}}^{+}\mathbf{E}\cdot \bar{\mathbf{v}}_n,
\end{equation}
with $\hat{\mathrm{S}}^{+}$ denoting the \textit{pseudoinverse} of $ \hat{\mathrm{S}}$. A standard inversion is infeasible, since LR theory imposes the condition
\begin{equation}
    \sum_{n}\int \frac{d\mathbf{p}}{(2 \pi)^2}\delta f_{\mathbf{p},n}=0. \label{zero_condition}
\end{equation}
This reflects the fact that summing the distribution function over all degrees of freedom yields the number of particles, a condition already satisfied by $f_{\mathbf{p}}^{(\mathrm{eq})}$.Ultimately, this renders the system of BEs overdetermined.

\section{Results and Discussion}\label{sec:res}
To make the problem more analytically tractable and, at the same time, to better understand the roles of mass and Rashba anisotropies, we analyze these two cases separately: (i) $m_x \neq m_y$ with $\alpha_x = \alpha_y = \alpha$, and $m_x = m_y = m$ with $\alpha_x \neq \alpha_y$. Each analytical result obtained via the Kubo formalism, Eq. (\ref{Kubo_2}), is validated through a numerical check implemented with the BE, Eqs. (\ref{BE_solution}) and (\ref{numerical_BE}). We consider this dual evaluation essential, as anisotropic Fermi surfaces lead to nontrivial momentum integrations, often requiring systematic Taylor expansions in the anisotropy parameters within the Kubo formalism which grow the algebraic complexity.
\subsection{Review of the CBA} 
In what follows,  we estimate the spin susceptibility within the CBA, wherein the GFs self-energy is replaced by a constant, $\Sigma^{R(A)}\rightarrow-\mathrm{i}/2 \tilde{\tau}$,
with $\tilde{\tau}$ having the units of time, and the vertex corrections are ruled out, requiring that  $\tilde{J}_{\beta}\rightarrow j_{\beta}$ in Eq. (\ref{Kubo_2}).
Considering, for simplicity, an electric field applied along the $x$-axis, the current operator is
\begin{equation}
    j_x=\frac{p_x}{m_x}-\alpha s_y,
\end{equation}
in case (i), and
\begin{equation}
    j_x=\frac{p_x}{m}-\alpha_x s_y,
\end{equation}
in case (ii).

For computational purposes, it is convenient to decompose the GFs into their spin components, $\mathcal{G}^{\mathrm{R(A)}}_i=(1/2)\mathrm{tr}\left[ \mathcal{G}^{\mathrm{R(A)}} s_i\right]$, with $i=0,x,y$, where each element is directly related to the projected GFs (PGFs), $\mathcal{G}^{\mathrm{R(A)}}_\pm$, on the two bands $\varepsilon_{+}$ and $\varepsilon_{-}$.
To show this, we express the Hamiltonian in the compact form 
\begin{equation}
    H=h_0 s_0 +\mathbf{h}\cdot\mathbf{s},
\end{equation}
where $h_0$ is the kinetic term, $\mathbf{s}$ is the vector of Pauli matrices, and $\mathbf{h} = (\alpha_y p_y,-\alpha_x p_x,0)$ is the effective magnetic field induced by the Rashba SOC. The two projectors associated to the helicity bands are $P_{\pm}=(s_0 \pm \hat{\mathrm{h}}\cdot \mathbf{s})/2$, allowing us to rewrite the GFs as
\begin{equation}
    \mathcal{G}^{\mathrm{R(A)}}=\left(\mathcal{G}^{\mathrm{R(A)}}_+ + \mathcal{G}^{\mathrm{R(A)}}_-\right)\frac{s_0}{2}+\left(\mathcal{G}^{\mathrm{R(A)}}_+ - \mathcal{G}^{\mathrm{R(A)}}_-\right)\frac{\hat{\mathrm{h}}\cdot \mathbf{s}}{2},
\end{equation}
where  $\hat{\mathrm{h}}=\mathbf{h}/|\mathbf{h}|$ and 
\begin{equation}
|\mathbf{h}|=\sqrt{(\alpha_xp_x)^2+(\alpha_yp_y)^2} \label{h_module}
\end{equation}
are the direction and module of the effective magnetic field, respecively.
The Kubo-Streda formula can then be expressed in terms of products of
 PGFs, $\sim\sum_{i,j \in \{-,+\}} \mathcal{G}^R_i \mathcal{G}^A_j$, where only the \textit{intra}-band terms, $\sim \mathcal{G}^R_{\pm} \mathcal{G}^A_{\pm}$, are retained. \textit{Inter}-band elements, which describe transitions between different bands at the same momentum and thus account for effects outside the Fermi surface, contribute to the response function only beyond leading order \cite{Rammer_2018}. What remains exists only within the Fermi surface, valid in the semiclassical regime, in accordance with Eq. (\ref{BE_ANG}). To see this explicitly, we can express products between  PGFs to the leading order in impurity concentration as
\begin{equation}
    \mathcal{G}^R_{\pm} \mathcal{G}^A_{\pm}=\mathrm{i}\tilde{\tau}\left(\mathcal{G}^R_{\pm}-\mathcal{G}^A_{\pm}\right)=-2\pi\tilde{\tau}\delta(\varepsilon_{\mathrm{F}}-\varepsilon_{\mathbf{p},n}),\label{delta}
\end{equation}
essentially playing the role of Eq. (\ref{RTA_sol}) for the BE. The calculation of the Kubo-Streda formula is now greatly simplified, as the delta function in Eq. (\ref{delta}) constraints the momentum integration within the Fermi surfaces, finally giving the result
\begin{equation}
    K^{(i)}_{yx}=\frac{m_y \alpha \tilde{\tau}}{2\pi n_i(1+\sqrt{\frac{m_y}{m_x})}}, \label{resp_CBA_m}
\end{equation}
in case (i), and 
\begin{equation}
K^{(ii)}_{yx}=\frac{\alpha_y m \tilde{\tau}}{2 \pi n_i(1+\frac{\alpha_y}{\alpha_x})}, \label{resp_CBA_a_2}
\end{equation}
in case (ii) in accordance with the literature \cite{Gaiardoni_2025}, where the relaxation time is promoted to a transport time $\tilde{\tau}\rightarrow 2 \tau_{\mathrm{tr}}$.


Eqs. (\ref{resp_CBA_m}) and (\ref{resp_CBA_a_2}) clearly show a dependency of the transverse spin response to the anisotropy parameters, controlled by the ratio between the masses and the Rashba couplings along the perpendicular axis. Apparently, tuning this ratio provides additional control over the response intensity compared to the isotropic case. However, the constant relaxation time $\tilde{\tau}$ is suspiciously independent of anisotropy, although, being proportional to the DOS, should reflect such dependency. Should this be the case, the responses behavior in Eqs. (\ref{resp_CBA_m}) and (\ref{resp_CBA_a_2}) could be radically different.
In the following sections, we will compute the spin susceptibility following the prescriptions in Sec. \ref{sec:method} and compare the outcomes with the results found in this section.

\begin{figure}
	\begin{centering}
		\includegraphics[scale=0.52]{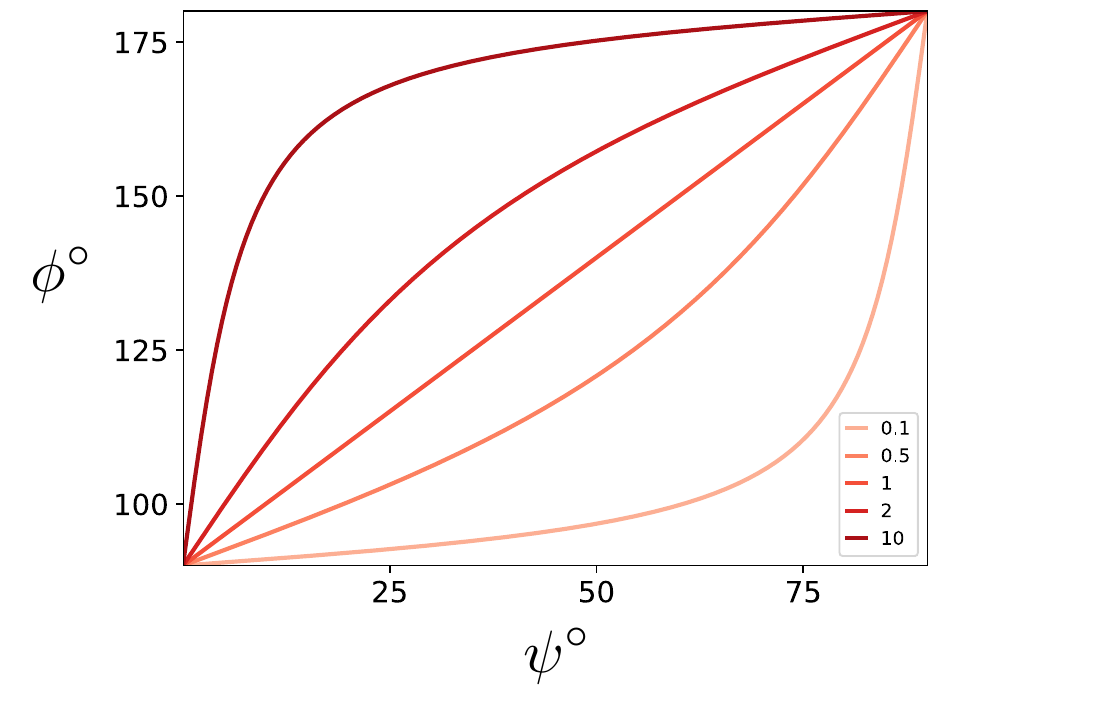}\caption{Inclination of the spin density as a function of the applied electric field angle $\psi$ and the ratio $\alpha_{\mathrm{r}}=\alpha_y/\alpha_x$ . Each curve corresponds to a different value of $\alpha_{\mathrm{r}}$, as indicated in the legend. The plots are computed numerically and are in perfect agreement with the analytical result given in Eq. (\ref{angle_law}).}\label{Rashba_angle}
		\par\end{centering}
\end{figure}

\subsection{Full inclusion of the vertex corrections beyond the CBA: mass anisotropy}
Focusing on mass anisotropy and resuming our last comment, we need to clarify the dependence of the self-energy on anisotropy. This is a relatively easy task, the result being
\begin{equation}
    \Sigma^{\mathrm{R(A)}} = \mp \frac{\mathrm{i}}{2}n_i u_0^2 \sqrt{m_x m_y}\, s_0,
\end{equation}
where only the anti-hermitian part of the self-energy has been retained. As expected intuitively, the relaxation time depends simultaneously on $m_x$ and $m_y$, as does the DOS, thereby invalidating the naive application of the CBA with a constant factor $\tilde{\tau}$. Our following goal is to evaluate the Kubo-Streda formula with the renormalized current vertex. 

Courtesy of the simple Hamiltonian structure, this task is feasible analytically with the help of a straightforward rearrangement of Eq. (\ref{BS_1}),
\begin{equation}
    \tilde{J}_x=\frac{p_x}{m_x}+\Gamma_x
\end{equation}
where 
\begin{equation}
\begin{split}
    \Gamma_x=-\alpha s_y+n_i u_0^2&\int\frac{d\mathbf{p}}{(2 \pi)^2}\mathcal{G}^{\mathrm{R}}_{\mathbf{p}}\frac{p_x}{m_x}\mathcal{G}^{\mathrm{A}}_{\mathbf{p}}\\&+n_i u_0^2\int\frac{d\mathbf{p}}{(2 \pi)^2}\mathcal{G}^{\mathrm{R}}_{\mathbf{p}}\Gamma_x\mathcal{G}^{\mathrm{A}}_{\mathbf{p}} \label{BSE_Gamma}
\end{split}
\end{equation}
is the momentum-independent part of the renormalized vertex entering the Bethe-Salpeter equation. Following the same decomposition scheme employed in Sec. \ref{sec:res}A, we find the second term on the right-hand side of Eq. (\ref{BSE_Gamma}) being equal to $\alpha s_y$ to the lowest order in the impurity concentration, exactly canceling the first term. Remarkably, this is allowed by a nontrivial compensation between the anisotropy carried by the self-energy and that contained in the DOS, which information is completely lost in the RTA. As a result, $\Gamma_x=0$, and the renormalized vertex reduces to
\begin{equation}
    \tilde{J}_x=\frac{p_x}{m_x},\label{bare_current}
\end{equation}
which is the bare vertex in the absence of Rashba coupling. This is a striking but well-known result, already established in the literature \cite{Raimondi_2005}, displaying an additional deviation from the RTA. Combining these results, the spin susceptibility becomes
\begin{equation}
    K_{yx}=\frac{\alpha}{2\pi n_i u_0^2},
\end{equation}
which coincides with the traditional EE \cite{Raimondi_2012}. In other words, mass anisotropy leaves the spin-current response unaffected at the semiclassical level, in contrast with current knowledge. The numerical approach based on QKT confirms this outcome, in agreement with our analytic analysis. 
On physical grounds, in retrospect, the above result is not surprising, given the fact the EE can be seen as the spin susceptibility response to the internal Rashba magnetic field induced by the distortion of the Fermi surface in the presence of an applied electric field.
This also provides another example, beyond the SHE in 2DEGs \cite{Raimondi_2005}, where vertex correction do not merely refine transport times but fundamentally change the transport \textit{phenomenology} of the studied system.

\subsection{Full inclusion of the vertex corrections beyond the CBA:  Rashba anisotropy} 

The considerations presented in the previous section can be reproduced here in full, especially the expression for the renormalized current vertex given in Eq.(\ref{BSE_Gamma}), with $\alpha\rightarrow\alpha_x$ and $m_x\rightarrow m$. In fact, once again we find that the second term on the right-hand side, which equals $\alpha_x s_y$, exactly cancels the first. As a result, the renormalized current reduces to the bare vertex in the absence of Rashba SOC, Eq. (\ref{bare_current}). This may seem surprising, because the GFs self-energy loses its anisotropic character, simply becoming $    \sim \mathrm{i}n_i u_0^2 m$, and therefore no longer compensates the anisotropy of the DOS. However, its role is taken over by the magnitude of the effective magnetic field $|\mathbf{h}|$, Eq.(\ref{h_module}), explicitly appearing in the off-diagonal part of the GFs and to which the DOS is proportional. 
Using these results in the Kubo formula, we finally find
\begin{equation}
    \mathbf{S}=\frac{\hat{\alpha}\,\mathbf{E}}{2\pi n_i u_0^2}, \label{spin_response_rashba}
\end{equation}
where
\begin{equation}
    \hat{\alpha}=\begin{pmatrix}
0 & -\alpha_y \\
\alpha_x & 0
\end{pmatrix}.
\end{equation}
The functional dependence of $\mathbf{S}$ from the physical parameters in play is the same as the conventional EE,  $\mathbf{S}^{\mathrm{iso}}$, 
which is explicitly recovered either when the electric field is applied along a principal axis or when Rashba isotropy is restored.
The distinction between the contribution of the different Rashba constants 
now becomes apparent, whereas it remained unseen in the isotropic case, with important consequences for the transport properties of the system. First, the magnitude of the spin density depends on the ratio between $\alpha_x$ and $\alpha_y$. For example, assuming $\alpha_x \neq0$ and an electric field applied  at an angle of $\pi/4$, we obtain $\mathrm{S}=\frac{\alpha_x E}{\sqrt{2}}\sqrt{1+(\alpha_y/\alpha_x)}$, which exceeds $\mathrm{S}^{\mathrm{iso}}$ if $\alpha_y>\alpha_x$ or remains lower otherwise. Second, the direction of the spin density is no longer, in general, perpendicular to the electric field, but follows the law
\begin{equation}
    \phi = \arctan \left(-\frac{\alpha_x}{\alpha_y} \cot \psi \right) + \pi, \label{angle_law}
\end{equation}
where $\phi = \arctan\left( \frac{S_y}{S_x}\right)$ and $\psi = \arctan\left( \frac{E_y}{E_x}\right)$. Clearly, if $\alpha_x=\alpha_y$, then $\phi = \psi +\pi/2$, that is the conventional EE, where the spin density is perpendicular to the applied electric field. 

The validity of Eq. (\ref{angle_law}) is tested numerically with the Boltzmann formalism, and depicted in Fig. (\ref{Rashba_angle}), wherin
we find a \textit{perfect match} with the analytical result. 
The plots in Fig. (\ref{Rashba_angle}) indicate that for small values of $\alpha_y/\alpha_x$ and $\psi$, the spin density is mainly transverse to the applied field, but as the ratio increases, it progressively aligns with the electric field. Conversely, for $\psi\rightarrow\pi/2$, the behavior is reversed. Beyond the fascinating deviation of this phenomenon from the conventional EE, the precise relation between the direction of the spin density and the ratio between the Rashba couplings can be exploited experimentally: a purely electrical excitation of the material indicates the degree of anisotropy of the Rashba coupling.

\begin{figure}
	\begin{centering}
		\includegraphics[scale=0.3]{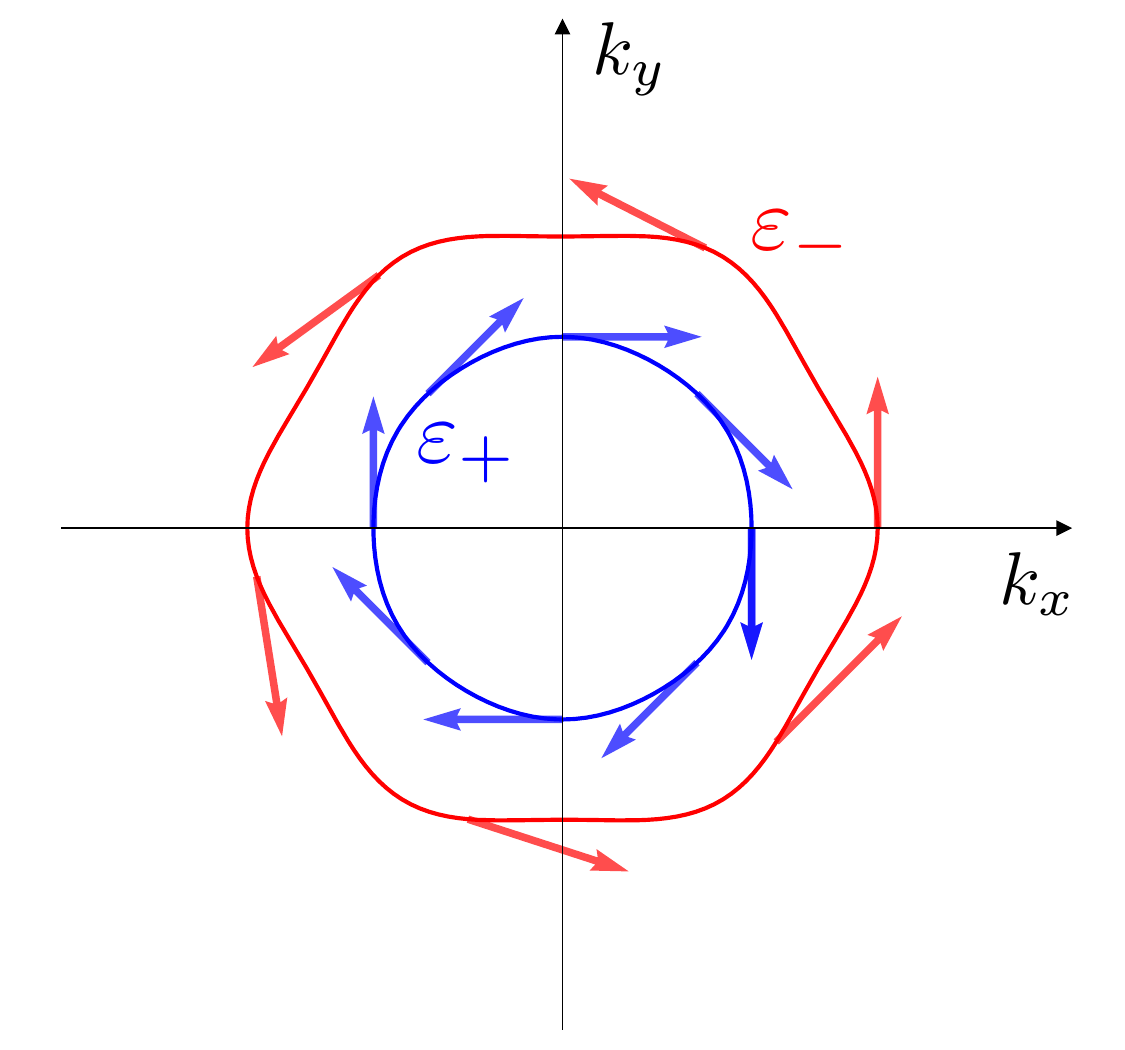}\caption{Fermi surfaces of a 2DEG with $C_{3v}$ symmetry in the presence of a warping term, Eq. (\ref{warping}). The figure is representative of Bi/Cu(1 1 1) systems \cite{Frantzeskakis_2011}, with $m=-0.29\,m_{\mathrm{e}}$, $\alpha=0.85\,\mathrm{eV}\,\text{\AA}$, $\lambda = 12\,\mathrm{eV}\,\text{\AA}^3$, and $\varepsilon_{\mathrm{f}}=-0.215\,\mathrm{eV}$. The Rashba splitting is $\sim0.1\,\mathrm{eV}.$}\label{fermi_disks_warp}
		\par\end{centering}
\end{figure}

\subsection{The warping term}

As promised in the Introduction, 
we extend our analysis to include an additional source of anisotropy in the low-energy effective Hamiltonian considered in this paper, Eq. (\ref{Ham}), which applies to a broad family of realistic materials of $C_{3v}$ symmetry \cite{Frantzeskakis_2011}, ranging from 2DEGs to surface states of three-dimensional topological insulators (TI).
Such symmetry rules out the anisotropies considered in the former sections of this work, imposing that $m=m_x=m_y$, $\alpha =\alpha_x=\alpha_y$. In this scenario, the SOC gives rise to the warping term \cite{fu2009}
\begin{equation}
H_w = \frac{\lambda}{2}(p_+^3+p_-^3)s_z,
    \label{warping}
\end{equation}
where $p_\pm =p_x\pm i p_y$, whose effect on the Fermi surfaces is illustrated in Fig.(\ref{fermi_disks_warp}). The eigenvalues read
\begin{equation}
\varepsilon_\pm (p_x,p_y) =\frac{p^2}{2m}\pm
\sqrt{\alpha^2p^2+\lambda^2p^6 \cos^2 (3 \theta)}
    \label{eigenvalues_warping}
\end{equation}
and, in contrast to systems with purely Rashba SOC (both isotropic and anisotropic), the eigenvectors in the presence of a warping term (\ref{warping}) acquire an out-of-plane spin component. 

\begin{figure}
	\begin{centering}
		\includegraphics[scale=0.45
    ]{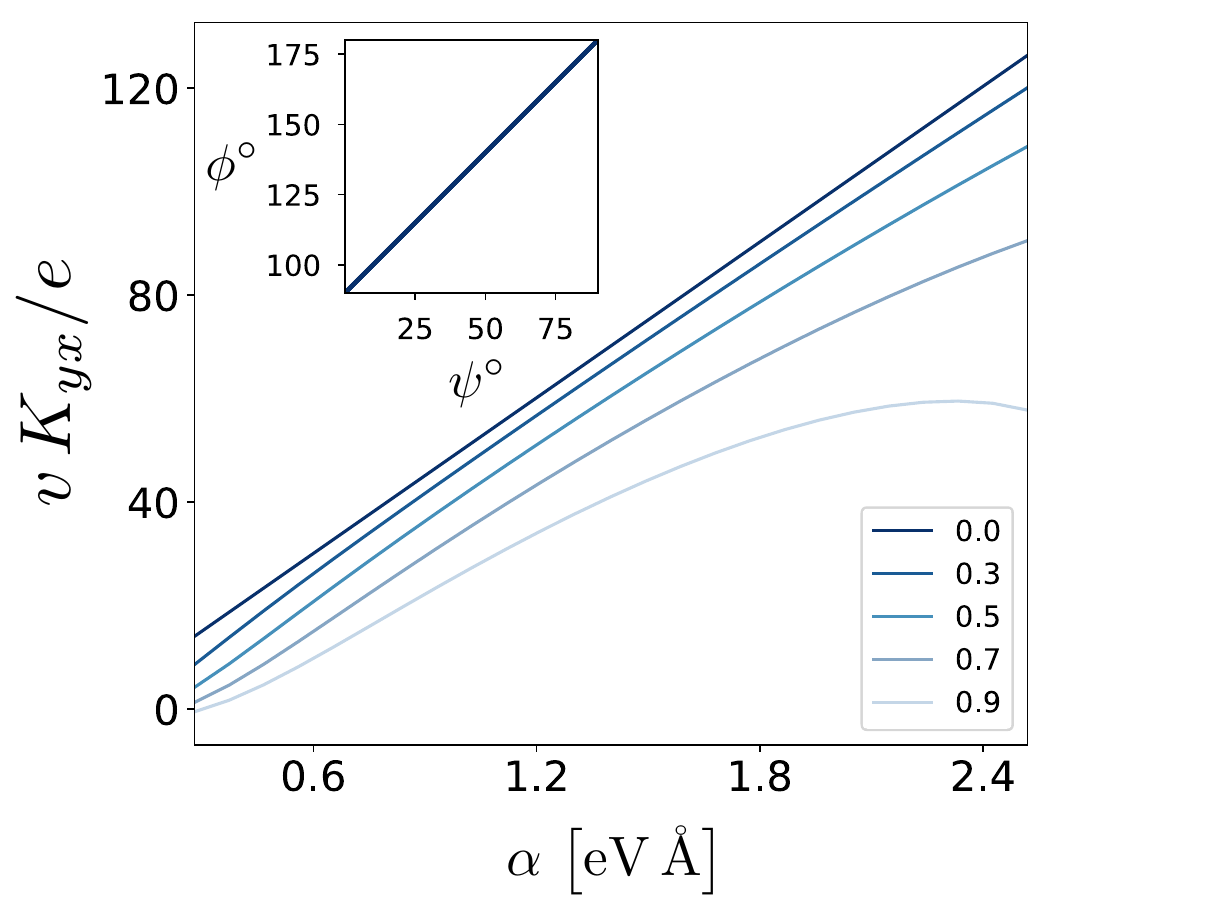}\caption{Rashba coupling dependence of the current-induced spin susceptibility for different values of the warping constant $\lambda$, listed in the legend. The inset illustrates the spin density inclination against the applied electric field angle, revealing perfect perpendicularity for any value of $\lambda$, alike the EE. The prototypical studied system is again Bi/Cu(1 1 1), introduced in Fig. (\ref{fermi_disks_warp}), where $v=5\times10^5\,\mathrm{m/s}$ is the Fermi velocity. The other parameters are $n_i=10^{16}\,\mathrm{m^{-2}}$ and $u_0=0.1\,\mathrm{eV}\,\mathrm{nm^2}.$}\label{warp_resp}
		\par\end{centering}
\end{figure}

An analytic solution to the transport problem is intractable in this system, and therefore a fully numerical approach is needed. The result is shown in Fig. (\ref{warp_resp}), which plots the predicted trend of the spin susceptibility as a function of the Rashba coupling strength and for different values of $\lambda$. In contrast to the Rashba anisotropy case, see Fig. (\ref{Rashba_angle}), the inclusion of a warping term in the Hamiltonian preserves the direction of the spin density perpendicular to the applied electric field, as illustrated in the inset. For this reason, and without loss of generality, the main plot shows the spin response component along the y-axis for an electric field applied along the x-axis. 
One effect of the warping anisotropy is the systematic decrease of the susceptibility as the warping strength increases, which becomes more pronounced for large values of the Rashba constant. This is largely attributed to the out-of-plane rotation of the electrons spin momentum at the Fermi surfaces and their departure from the ideal circular path, as pointed out in the literature \cite{Johansson_2016}. A second effect we report is a progressive deviation from the linear dependence of the susceptibility on $\alpha$, characteristic of the EE in 2DEGs, which further contributes to the decrease of the spin response.

\section{Conclusions}\label{sec:conc}

In this paper we had a fresh look at the influence of the lowering of the full rotational symmetry to the $C_{2v}$ and $C_{3v}$ symmetry groups on the ISGE (also known as Edelstein effect) in the presence of Rashba SOC.
At variance with previous investigations in the literature, we have established that vertex corrections are crucial and their subtle effect is not captured by constant broadening approximation. For instance, in the case of mass anisotropy, in contrast with previous reported results, we find that the ISGE remains unaffected. More, importantly, in the case of Rashba SOC anisotropy, the $C_{2v}$ reduced symmetry manifests in a different strength for the ISGE along the two axes of the two-dimensional electron gas. In particular, in general for arbitrary direction of the electric field, the induced spin polarization is no longer simply perpendicular to the electric field but makes an angle larger or smaller than $\pi/2$ depending on the anisotropy ratio of the Rashba couplings. In the case of $C_{3v}$ symmetry, the theoretical analysis has been a full numerical implementation of the general scheme provided by the QKT approach previously developed by one of the authors. In this respect our paper also shows, in a relatively simple Hamiltonian case, the feasibility of a general method to treat disorder effects beyond the often used constant broadening approximation.

\begin{acknowledgments}
TV and RR  thank  Mairbek Chshiev,  Jing Li, Libor Voj\'a\v{c}ek for fruitful discussions. One of the authors (TV) acknowledges support by the EIC Pathfinder OPEN Grant No. 101129641 “OBELIX” and a France 2030 government grant managed by the French National Research Agency PEPR SPIN Grant No. ANR-22-EXSP0009 (SPINTHEORY). 
\end{acknowledgments}

\bibliography{BIB}
\end{document}